\documentclass[pre,twocolumn,floatfix]{revtex4}
\usepackage{graphicx,amsmath,amssymb,mathptm,multirow}
\begin{document}
\title{Scale-free random branching tree in supercritical phase}
\author{D.-S. Lee,$^1$ J. S. Kim,$^2$ B. Kahng,$^{1,2}$ and D. Kim$^2$}
\affiliation{{$^1$ Center for Complex Network Research and
Department of Physics, University of Notre Dame,
Notre Dame, IN 46556, USA}\\
{$^2$ CTP {\rm \&} FPRD, School of Physics and Astronomy, Seoul
National University, NS50, Seoul 151-747, Korea}}
\date{\today}
\begin{abstract}
We study the size and the lifetime distributions of scale-free
random branching tree in which $k$ branches are generated from a
node at each time step with probability $q_k\sim k^{-\gamma}$. In
particular, we focus on finite-size trees in a supercritical phase,
where the mean branching number $C=\sum_k k q_k$ is larger than 1.
The tree-size distribution $p(s)$ exhibits a crossover behavior when
$2 < \gamma < 3$; A characteristic tree size $s_c$ exists such that
for $s \ll s_c$, $p(s)\sim s^{-\gamma/(\gamma-1)}$ and for $s \gg
s_c$, $p(s)\sim s^{-3/2}\exp(-s/s_c)$, where $s_c$ scales as $\sim
(C-1)^{-(\gamma-1)/(\gamma-2)}$. For $\gamma
> 3$, it follows the conventional mean-field solution, $p(s)\sim
s^{-3/2}\exp(-s/s_c)$ with $s_c\sim (C-1)^{-2}$. The lifetime
distribution is also derived. It behaves as $\ell(t)\sim
t^{-(\gamma-1)/(\gamma-2)}$ for $2 < \gamma < 3$, and $\sim t^{-2}$
for $\gamma > 3$ when branching step $t \ll t_c \sim (C-1)^{-1}$,
and $\ell(t)\sim \exp(-t/t_c)$ for all $\gamma > 2$ when $t \gg
t_c$. The analytic solutions are corroborated by numerical results.
\end{abstract}
\pacs{02.50.-r, 05.40.-a, 89.75.Da}

\maketitle
\section{Introduction}
A tree is a graph with no loop within it. Owing to the simplicity of
its structure and amenability of analytic studies, tree graph has
drawn considerable attentions in many disciplines of scientific
researches. Scale-free (SF) random branching tree, in which the
number of branches $k$ generated from a node is stochastic following
a power-law distribution, $q_k\sim k^{-\gamma}$, is particularly
interesting here. Such trees can be found in various phenomena such
as the trajectories of cascading failure in the sandpile model on SF
networks~\cite{sandpile}, epidemic spreading on SF
networks~\cite{epidemic0,epidemic1}, aftershock propagation in
earthquake~\cite{earthquake0,earthquake1}, random spanning tree or
skeleton of SF networks~\cite{fractal}, phylogenetic
tree~\cite{taxanomy}, etc. Here, SF network is the network with the
degree distribution following a power law $P_d(k)\sim
k^{-\lambda}$~\cite{sf0,sf1,sf2}. So far, several analytic studies
have been performed to understand structural properties of SF
branching trees~\cite{review}. However, most works are focused on
the critical case, where the mean branching number $C \equiv\sum_k k
q_k$ is equal to 1, motivated by universal feature of scale
invariance observed in nature and society.

Recent studies, however, show that the structure of real-world
networks may have been designed upon supercritical
trees~\cite{fractal}. Supercritical trees, where the mean branching
number $C >1$, turn out to act as a skeleton of some fractal
networks such as the world-wide web. Here skeleton~\cite{skeleton}
is defined as a spanning tree formed by edges with highest
betweenness centrality or loads~\cite{bc,load}. A supercritical
branching tree can grow indefinitely with a nonzero probability,
which is the most marked difference from critical ($C=1$) or
subcritical ($C<1$) tree that cannot grow infinitely. Moreover, the
total number of offsprings $s(t)$ generated from a single root
(ancestor) up to a given generation $t$ can increase exponentially
in supercritical trees and this is reminiscent of the small-world
behavior: The mean distance between nodes scales logarithmically as
a function of the total number of nodes~\cite{review}.

Due to the mean branching number being larger than 1, some
supercritical trees may be alive in a very long time limit. The
tree-size distribution of those surviving trees in the supercritical
phase has been derived in the mean-field framework~\cite{rios},
which follows a power law, $p(s)\sim s^{-2}$. Here, we consider
finite-size trees in the supercritical phase. In spite of the large
mean branching number, some trees do not grow infinitely even in the
supercritical phase. For such finite-size trees in the supercritical
phase, we derive the tree-size and the lifetime distributions using
the generating function technique~\cite{harris}. Distinguished from
the critical case, the generating function of the tree-size
distribution exhibits two singular behaviors in the supercritical
phase and thereby a crossover behavior of the tree-size distribution
can arise when $2 < \gamma < 3$. We present in detail the derivation
of all these analytic solutions in the following sections. The
tree-size and lifetime distributions predicted by analytic solutions
are confirmed by numerical simulations. This is important in itself
for understanding the branching trees whose structure changes
drastically depending on the phase. Since the branching tree
approach can be applied to numerous systems, our results should be
useful for future diverse applications as well.

\section{Tree-size distribution}

Let us consider the branching process that each node generates $k$
offsprings with probability $q_k$,
\begin{equation}
q_k =\left\{
\begin{array}{ll}
1-\frac{C\,\zeta(\gamma)}{\zeta(\gamma-1)}&~~~ {\rm for}~~  k=0,
\\[2.5mm]
\frac{C}{\zeta(\gamma-1)} k^{-\gamma} &~~~ {\rm for}~~ k\geq 1,
\end{array} \right.
\label{eq:qk}
\end{equation}
where $C$ is constant in the range of $0 < C
<\zeta(\gamma-1)/\zeta(\gamma)$ with the Riemann-zeta function
$\zeta(x)$, and $\gamma$ is larger than $2$, ensuring that
$\zeta(\gamma-1)$ is finite. Then, $C$ is automatically identical to
the mean branching number, i.e. the average number of offsprings
$C=\sum_{k=0}^{\infty} kq_k$ generated from a node.
When $C < 1$, the number of offsprings decreases on average as
branching proceeds and it vanishes eventually. Thus, branching tree
has finite lifetime with probability one. When $C >1$, as branching
proceeds, the number of offsprings can increase exponentially with
non-zero probability. The case of $C=1$ is marginal: Offsprings
persist, neither disappear nor flourish on average. A branching tree
generated through the stochastic process (1) is a SF branching tree,
because its degree distribution follows a power law, $P_d(k_d)\sim
k_d^{-\gamma}$ asymptotically. Degree $k_d$ of each node in the tree
is related to the branching number $k$ of that node as $k_d=k+1$ but
for the root, $k_d=k$.

\subsection{Generating function method}

A tree grows as each of the youngest nodes generates their
offsprings following the probability $q_k$ in Eq.~(\ref{eq:qk}).
This evolution is regarded as a process in a unit time step. When a
node generates no offspring with probability $q_0$, it remains
inactive in further time steps. We define $p_t(s)$ as the fraction
of trees with total number of nodes $s$ at time $t$. By definition,
$p_0(s)=\delta_{s,1}$. Then, $p_{t+1}(s)$ can be written in terms of
$p_t(s)$ as
\begin{equation}
p_{t+1}(s) = \sum_{k=0}^\infty q_k \sum_{s_1,s_2,\ldots,s_k} p_t(s_1)
  p_t(s_2) \cdots p_t(s_k) \delta_{\sum_{i=1}^k s_i, s-1}.
\label{p_relation}
\end{equation}
Defining the generating functions, $\mathcal{Q}(\omega)
=\sum_{k=0}^\infty q_k \omega^k$ and $\mathcal{P}_{t} (y)
=\sum_{s=1}^\infty p_t(s) y^s$, and applying them to
(\ref{p_relation}), one can obtain that
\begin{equation}
\mathcal{P}_{t+1}(y)=y \mathcal{Q}(\mathcal{P}_t(y)).
\label{g_relation}
\end{equation}
Let us consider the tree-size distribution in the $t\to \infty$
limit, i.e., $p(s)=\lim_{t\to \infty} p_t (s)$ and its generating
function $\mathcal{P}(y)=\lim_{t\to \infty} \mathcal{P}_t(y)$.
However, some trees may grow infinitely in the supercritical phase,
which makes $\mathcal{P}(y)=\sum_s p(s) y^s$ ill-defined at $y=1$.
So we limit the summation in $\mathcal{P}(y)$ over finite trees
only, i.e., ${\mathcal{P}}(y)=\sum_{{\rm finite}~s}p(s)y^s$. This is
equivalent to defining ${\cal P}(1)=\lim_{y \to 1}{\mathcal
{P}}(y)$. Then, Eq.(\ref{g_relation}) gives the relation in the
$t\to \infty$ limit,
\begin{equation}
\mathcal{P}(y) = y \mathcal{Q}(\mathcal{P}(y)).
\label{eq:gen}
\end{equation}
The next step is to extract a singular part of $\mathcal{P}(y)$ from
Eq.~(\ref{eq:gen}), and then to derive the behavior of $p(s)$ for
$s\gg 1$.

The power-law form of $q_k$ in Eq.~(\ref{eq:qk}) results in the
expansion of $\mathcal{Q}(\omega)$ around $\omega=1$:
\begin{eqnarray}
&&\mathcal{Q}(\omega) = 1 - C (1-\omega) + \frac{B(\gamma)}{2}(1-\omega)^2
+\cdots
\nonumber\\
&&+ \left\{
\begin{array}{ll}
A(\gamma)(1-\omega)^{\gamma-1} & (\gamma\ne{\rm integer})\\[2.5mm]
\frac{(-1)^{\gamma}}{\Gamma(\gamma)} (1-\omega)^{\gamma-1} \ln (1-\omega) & (\gamma={\rm integer})
\end{array}
\right. +\cdots, \label{eq:qw}
\end{eqnarray}
where $B(\gamma)= C [\zeta(\gamma-2)/\zeta(\gamma-1)-1]$, and
$A(\gamma) = C \Gamma(1-\gamma)/\zeta(\gamma-1)$ with the Gamma
function $\Gamma(x)$. The inverse function
$y=\mathcal{P}^{-1}(\omega)$ is then expanded as
\begin{eqnarray}
y&=&\mathcal{P}^{-1}(\omega)=\frac{\omega}{\mathcal{Q}(\omega)}\sim 1+\Delta(1-\omega) - \frac{B(\gamma)}{2}  (1-\omega)^2
+ \cdots \nonumber\\
&-&\left\{\begin{array}{ll}
A(\gamma)(1-\omega)^{\gamma-1} & (\gamma\ne{\rm integer})\\[2.5mm]
\frac{(-1)^{\gamma}}{\Gamma(\gamma)} (1-\omega)^{\gamma-1} \ln (1-\omega) & (\gamma={\rm integer})
\end{array}
\right. + \cdots,
\label{eq:expand}
\end{eqnarray}
where $\Delta \equiv C-1$. We recall that $\Delta$ is positive
(negative) in the supercritical (subcritical) regime and $0$ in the
critical case. Here we focus on the supercritical case of $\Delta >
0$ and being very small, but the obtained result can be naturally
extended to large-$\Delta$ cases.

\begin{figure}[t]
\includegraphics[width=8cm]{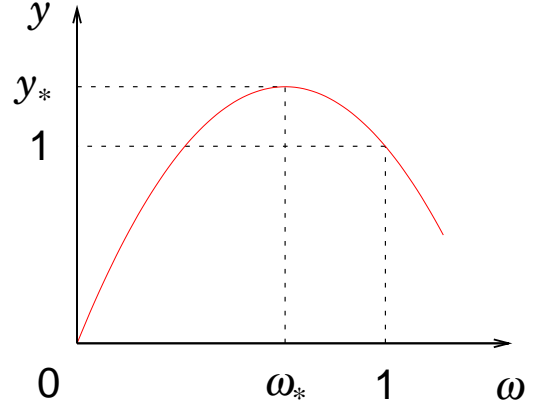}
\caption{(Color online) Schematic plot of the function
$y=\omega/{\cal Q}(\omega)$ in the supercritical phase. The
$dy/d\omega=0$ occurs at $\omega=\omega_* < 1$. } \label{fig:y}
\end{figure}

\subsection{The singularity at $y=y_* > 1$}
Let us investigate how $y$ behaves as $\omega$ decreases
from $1$ to $0$. For $\Delta>0$, as $\omega$ decreases from $1$ to
$\omega_*$, $y$ increases from $1$ to $y_*$ and then decreases to
zero as shown in Fig.\ref{fig:y}, where $\omega_*$ satisfying
$(d/d\omega)[\omega /\mathcal{Q}(\omega)]|_{\omega=\omega_*}=0$
locates less than 1. This feature is distinguished from the solution
$\omega_*=1$ for the critical case. It is obtained that $\omega_*$
depends on $\Delta$ as
\begin{equation}
1-\omega_* \equiv \epsilon_* \sim \left\{
\begin{array}{ll}
\Delta & {\rm for}~~~ \gamma>3,\\[2.5mm]
\Delta /\ln (1/\Delta) & {\rm for}~~~ \gamma=3,\\[2.5mm]
\Delta^{1/(\gamma-2)} & {\rm for}~~~ 2< \gamma <3.
\end{array}\right.
\label{eq:omega}
\end{equation}
The value $y_*$, determined by the relation
$y_*=\omega_*/\mathcal{Q}(\omega_*)$, locates at
\begin{equation}
y_*-1\equiv \delta_* \sim \left\{
\begin{array}{ll}
\Delta^2 & {\rm for}~~~ \gamma>3,\\[2.5mm]
\Delta^2 /\ln (1/\Delta) & {\rm for}~~~ \gamma=3,\\[2.5mm]
\Delta^{(\gamma-1)/(\gamma-2)} & {\rm for}~~~ 2< \gamma <3.
\end{array}\right.
\label{eq:y}
\end{equation}
The curve $y=\omega/\mathcal{Q}(\omega)$ in the region
$\omega>\omega_*$ is just the analytic continuation of the inverse
function $y=\mathcal{P}^{-1}(\omega)$ that is analytic for
$\omega<\omega_*$~\cite{otter}.

The right-hand-side of Eq.~(\ref{eq:expand}) for $\omega<\omega_*$
is expanded around $\omega_*$ as
\begin{equation}
y\simeq y_* +\sum_{n=2}^\infty \frac{D_n(\gamma)}{n!}(\omega_*-\omega)^n,
\label{eq:expand1}
\end{equation}
when $\omega$ is close to $\omega_*$ such that
\begin{equation}
\max_{n\geq
2}\frac{D_{n+1}(\gamma)}{D_{n}(\gamma)(n+1)}(\omega_*-\omega) \ll
1.\label{eq:cond1}
\end{equation}
Here $D_n(\gamma)$ is the $n$th derivative of
$\omega/\mathcal{Q}(\omega)$ at $\omega_*$. For $n=2$,
\begin{equation}
D_2(\gamma)\sim \left\{
\begin{array}{ll}
-B(\gamma) & {\rm for}~~~ \gamma>3,\\[2.5mm]
\ln \Delta & {\rm for}~~~ \gamma=3,\\[2.5mm]
-\Delta^{(\gamma-3)/(\gamma-2)} & {\rm for}~~~ 2< \gamma <3.
\end{array}\right.
\label{eq:d2}
\end{equation}
This result is used for future discussions. Keeping only the
quadratic term $(\omega_*-\omega)^2$ in Eq.(\ref{eq:expand1}), one
obtains the leading singular behavior of ${\mathcal P}(y)$ at $y_*$,
\begin{equation} \omega=\mathcal{P}(y)\sim \omega_* -
\sqrt{\frac{2(y_*-y)}{|D_2(\gamma)|}}. \label{eq:py1}
\end{equation}
In fact such a square-root singularity at $y=y_*$ is generic
regardless of the form of the branching probability when
$q_0+q_1<1$~\cite{otter}, yielding the asymptotic behavior of $p(s)$
given by
\begin{equation}
p(s)\sim  b(\Delta)s^{-3/2} \exp(-s/s_*), \label{eq:ps1}
\end{equation}
where the coefficient $b(\Delta)\sim
\Delta^{-(\gamma-3)/[2(\gamma-2)]}$ for $2 < \gamma < 3$,
$1/\sqrt{\ln (1/\Delta)}$ for $\gamma=3$ and constant for $\gamma
> 3$, and $s_* = (\ln y_*)^{-1}$.
\subsection{The singularity at $y=1$}

When $\omega$ is far from $\omega_*$ such that the linear term with
the coefficient $\Delta$ is not comparable to the next-order term,
another singularity becomes dominant. The next-order term is the
quadratic term for $\gamma>3$ and the non-analytic term for
$2<\gamma\leq 3$. To be precise, if the condition, $1-\omega\gg
\Delta$ for $\gamma>3$, $-(1-\omega)\ln (1-\omega)\gg \Delta$ for
$\gamma=3$, and $1-\omega\gg \Delta^{1/(\gamma-2)}$ for
$2<\gamma<3$, holds, then the linear term is negligible compared
with the next order terms, and then Eq.(\ref{eq:expand}) is reduced
to
\begin{equation}
y \sim 1 - \left\{
\begin{array}{ll}
\frac{B(\gamma)}{2}(1-\omega)^2 & {\rm for}~~~ \gamma>3,\\[2.5mm]
-\frac{1}{2} (1-\omega)^2 \ln (1-\omega) & {\rm for}~~~
\gamma=3,\\[2.5mm]
A(\gamma)(1-\omega)^{\gamma-1} & {\rm for}~~~ 2<\gamma<3.
\end{array}\right.
\label{eq:expand2}
\end{equation}
The generating function $\mathcal{P}(y)$ then behaves as
\begin{equation}
\omega=\mathcal{P}(y) \sim 1 - \left\{
\begin{array}{ll}
\sqrt{\frac{2(1-y)}{B(\gamma)}} & {\rm for}~~~ \gamma>3,\\[2.5mm]
\sqrt{\frac{4(1-y)}{|\ln (1-y)|}} & {\rm for}~~~ \gamma=3,\\[2.5mm]
\left(\frac{1-y}{A(\gamma)}\right)^{1/(\gamma-1)} & {\rm for}~~~
2<\gamma<3.
\end{array}\right.
\label{eq:py2}
\end{equation}
From this result, one can obtain the tree-size distribution as
\begin{equation}
p(s) \sim \left\{
\begin{array}{ll}
s^{-3/2} & {\rm for}~~~ \gamma>3,\\[2.5mm]
s^{-3/2}(\ln s)^{-1/2} & {\rm for}~~~ \gamma=3,\\[2.5mm]
s^{-\gamma/(\gamma-1)} & {\rm for}~~~ 2<\gamma<3.
\end{array}\right.
\label{eq:ps2}
\end{equation}

\subsection{Crossover behavior between the two singularities}

The two singular behaviors of $\mathcal{P}(y)$ in the forms of
Eqs.~(\ref{eq:py1}) and (\ref{eq:py2}) occurring at $y=y_*$ and
$y=1$, respectively, enables us to determine the ranges of size $s$
where the formulae of Eqs.(\ref{eq:ps1}) and (\ref{eq:ps2}) are
valid. In particular, when $2<\gamma\leq 3$, the asymptotic
behaviors in Eqs.(\ref{eq:ps1}) and (\ref{eq:ps2}) differs from each
other and thus there should be a crossover behavior in the tree-size
distribution.

\begin{figure}[t]
\includegraphics[width=8cm]{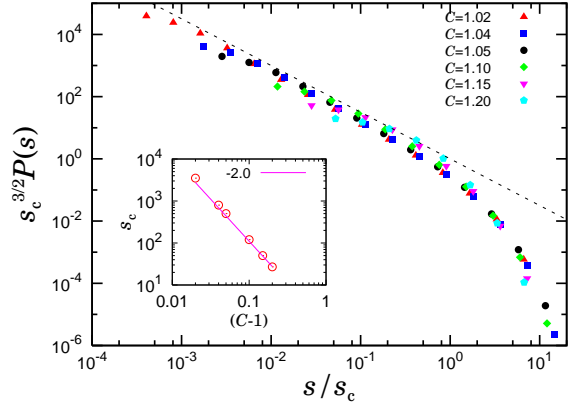}
\caption{(Color online) The tree-size distribution $p(s)$ for
$\gamma=3.3$ for various values of $C$ in the scaling form
Eq.~(\ref{eq:ps_gamma>3}). Dashed line is guideline with a slope of
$-3/2$. Inset: Dependence of the characteristic size $s_c$ on the
mean branching number $C$.} \label{fig:ps35}
\end{figure}

The ranges of $\omega$ in which Eqs.~(\ref{eq:expand1}) and
(\ref{eq:expand2}) are valid are closely related to those of $y$ for
Eqs.~(\ref{eq:py1}) and (\ref{eq:py2}) and that of $s$ for
Eqs.~(\ref{eq:ps1}) and (\ref{eq:ps2}), respectively. Here we find
those ranges of $\omega$, $y$, and $s$, and then determine the
crossover in the tree-size distribution $p(s)$.

First, we study valid ranges of Eqs.~(\ref{eq:expand1}),
(\ref{eq:py1}), and (\ref{eq:ps1}). The coefficient $D_n(\gamma)$ in
Eq.~(\ref{eq:expand1}) behaves as $\sim (1-\omega_*)^{\gamma-1-n}$
for $n>\gamma-1$ due to the non-analytic term
$(1-\omega)^{\gamma-1}$ in Eq.~(\ref{eq:expand}) when $\gamma$ is
not integer. Then, it follows that $[D_n(\gamma)
(n+1)/D_{n+1}(\gamma)]\sim 1/(1-\omega_*) \equiv 1/\epsilon_*$.
Thus, the condition (\ref{eq:cond1}) can be rewritten as
$\omega_*-\epsilon_c^{>}\ll \omega<\omega_*$, where
$\epsilon_c^{>}\sim \Delta$ for $\gamma>3$,
$\epsilon_c^>\sim\Delta/\ln(1/\Delta)$ for $\gamma=3$, and
$\epsilon_c^>\sim \Delta^{1/(\gamma-2)}$ for $2<\gamma<3$ from
Eq.(\ref{eq:omega}). The corresponding range of $y$ is
$y_*-\delta_c^{>}\ll y<y_*$, where $\delta_c^{>}$ is given by $\sim
\Delta^2$ for $\gamma>3$, $\sim \Delta^2/\ln(1/\Delta)$ for
$\gamma=3$, and $\sim \Delta^{(\gamma-1)/(\gamma-2)}$ for
$2<\gamma<3$ by using Eqs.(\ref{eq:y}) and (\ref{eq:py1}).

To find valid range of $s$ for $p(s)$ in Eq.(\ref{eq:ps1}), we use
the fact that the singular functional behavior of $\mathcal{P}(y)$
around $y=\tilde{y}$ is determined by that of $p(s)$ around
$s=\tilde{s}$, where $\tilde y$ and $\tilde s$ are related as
$\tilde{y}^{\tilde{s}}\sim 1$. Then, one can find that $s_c^{>}=|\ln
(y_*-\delta_c^>)|^{-1}\sim (\delta_*-\delta_c^{>})^{-1}$, so that
$s_c^{>}\sim \Delta^{-2}$ for $\gamma>3$, $\Delta^{-2}\ln
(1/\Delta)$ for $\gamma=3$, and $\Delta^{-(\gamma-1)/(\gamma-2)}$
for $2<\gamma<3$. For the range $s \gg s_c^>$, the formula
(\ref{eq:ps1}) is valid.

Second, we check the validities of Eqs.~(\ref{eq:expand2}),
(\ref{eq:py2}), and (\ref{eq:ps2}). Comparing the magnitude of the
linear term and the next-order term in Eq.~(\ref{eq:expand}), we
find that Eq.~(\ref{eq:expand2}) is valid for $\omega\ll
1-\epsilon_c^<$, where $\epsilon_c$ behaves as $\Delta$ for
$\gamma>3$, $\Delta/\ln(1/\Delta)$ for $\gamma=3$, and
$\Delta^{1/(\gamma-2)}$ for $2<\gamma<3$. The corresponding range of
$y$ for Eq.~(\ref{eq:py2}) is given as $y\ll 1-\delta_c^<$, where
$\delta_c^<\sim \Delta^2$ for $\gamma>3$, $\delta_c^<\sim
\Delta^2/\ln(1/\Delta)$ for $\gamma=3$, and $\delta_c^<\sim
\Delta^{(\gamma-1)/(\gamma-2)}$ for $2<\gamma<3$. The corresponding
range of $s$ for Eq.~(\ref{eq:ps2}) is $s\ll s_c^<$ with
$s_c^{<}=|\ln (1-\delta_c^<)|^{-1}\sim (\delta_c^{<})^{-1}$ given by
$s_c^{<}\sim \Delta^{-2}$ for $\gamma<3$, $s_c^{<}\sim
\Delta^{-2}\ln (1/\Delta)$ for $\gamma=3$, and $s_c^{<}\sim
\Delta^{-(\gamma-1)/(\gamma-2)}$ for $2<\gamma<3$.

As already noticed, the crossover sizes $s_c^>$, $s_c^<$, and $s_*$
are consistent for all values of $\gamma$ within
$\Delta$-dependence, and thereby, we use the notation $s_c$ for all
of them. The overall behavior of the tree-size distribution is
obtained by combining Eqs.~(\ref{eq:ps1}) and (\ref{eq:ps2}). For
$\gamma>3$, there is no need to introduce a crossover. Thus, it
leads to
\begin{equation}
p(s)\sim s^{-3/2} \exp(-s/s_c) \quad (\gamma>3),
\label{eq:ps_gamma>3}
\end{equation}
for all $s$. And $s_c\sim \Delta^{-2}$. As $\Delta$ increases, the
cut-off $s_c$ decreases and the exponential-decaying pattern
prevails.

\begin{figure}[t]
\includegraphics[width=8cm]{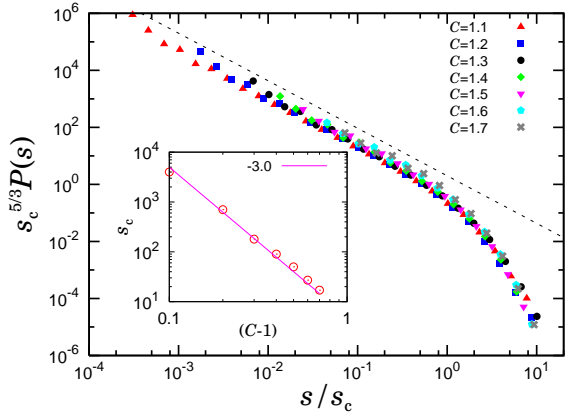}
\caption{(Color online) The tree-size distribution $p(s)$ for
$\gamma=2.5$ for various values of $C$ in the scaling form
Eq.~(\ref{eq:ps_2<gamma<3}). Dashed line is guideline with a slope
of $-5/3$. Inset: Dependence of the characteristic size $s_c$ on the
mean branching number $C$.} \label{fig:ps25}
\end{figure}

When $\gamma=3$, $p(s)$ is given by
\begin{equation}
p(s)\sim \left\{
\begin{array}{ll}
s^{-3/2}(\ln s)^{-1/2} & {\rm for} \ s\ll s_c, \\[2.5mm]
s^{-3/2} \exp(-s/s_c) & {\rm for} \ s\gg s_c,
\end{array}\right.
\quad (\gamma=3)
\label{eq:ps_gamma=3}
\end{equation}
where $s_c\sim \Delta^{-2}\ln(1/\Delta)$. Similarly, for
$2<\gamma<3$, we find that
\begin{equation}
p(s)\sim \left\{
\begin{array}{ll}
s^{-\gamma/(\gamma-1)} & {\rm for} \ s\ll s_c, \\[2.5mm]
s^{-3/2}\exp(-s/s_c) & {\rm for} \ s\gg s_c,
\end{array}\right.
\quad (2<\gamma<3)
\label{eq:ps_2<gamma<3}
\end{equation}
where both $s_c \sim \Delta^{-(\gamma-1)/(\gamma-2)}$. As $\Delta\to
0$, $s_c$ diverges, and the power-law behavior prevails.

We invoke numerical simulations to confirm our analytic solutions.
Figs.~\ref{fig:ps35} and \ref{fig:ps25} show the tree size
distributions for $\gamma=3.3$ and $\gamma=2.5$ in the scaling
forms, Eq.~(\ref{eq:ps_gamma>3}) and Eq.~(\ref{eq:ps_2<gamma<3}),
respectively. The data are well collapsed into the predicted
formulas for different $C$ values for both cases.

\section{Lifetime distribution}

\begin{figure}[t]
\includegraphics[width=8cm]{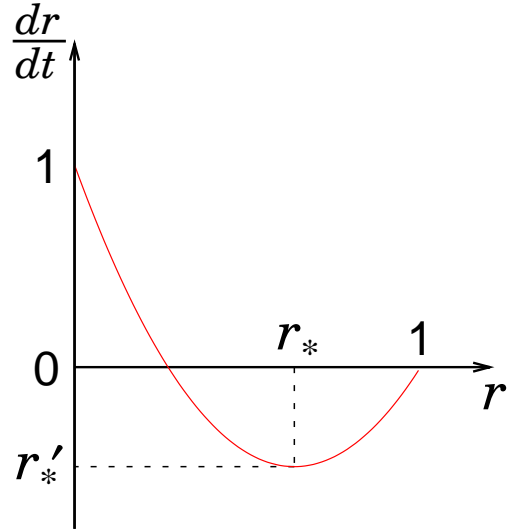}
\caption{(Color online) Schematic plot of the function $dr/dt={\cal
Q}(r)-r$ in the supercritical phase. $(\partial/\partial
r)(dr/dt)=0$ occurs at $r_*$, at which $dr/dt$ is denoted as
$r_*^{\prime}$.} \label{fig:lifetime}
\end{figure}

Next we solve the lifetime distribution $\ell(t)$. This is defined
as the probability that the branching process stops at $t$. To
derive $\ell(t)$, we first introduce the probability that the
branching process stops at or prior to time $t$, denoted by $r(t)$.
Then $\ell(t)$ is given as $\ell(t)=r(t+1)-r(t)$. The probability
distribution $r(t)$ is related to $r(t-1)$ as
\begin{equation}
r(t) = \sum_{k=0}^\infty q_k  [r(t-1)]^k = \mathcal{Q}(r(t-1)).
\end{equation}
Thus, we are given approximately a differential equation for $r(t)$,
\begin{equation}
\frac{dr(t)}{dt}\approx \ell(t)=\mathcal{Q}(r(t)) - r(t),
\label{eq:ltgen}
\end{equation}
Expanding the right hand side of Eq.~(\ref{eq:ltgen}) around $r=1$,
one can see its asymptotic behavior. Using Eq.~(\ref{eq:qw}) again,
we find $dr/dt$ in the long time limit as follows:
\begin{eqnarray}
\frac{dr}{dt}&=&\mathcal{Q}(r)-r = - \Delta (1-r) +
\frac{B(\gamma)}{2}(1-r)^2 +\cdots
\nonumber\\
&+&\left\{
\begin{array}{ll}
A(\gamma)(1-r)^{\gamma-1} & (\gamma\ne{\rm integer})\\[2.5mm]
\frac{(-1)^{\gamma}}{\Gamma(\gamma)} (1-r)^{\gamma-1} \ln (1-r) &
(\gamma={\rm integer})
\end{array}
\right. +\cdots.
  \label{eq:qr}
\end{eqnarray}
What we can see in this relation is that the value of $r^{\prime}$
is zero at $r=1$. It decreases as $r$ decreases until it reaches
$r_*$ where $(d/dr)[\mathcal{Q}(r)-r]|_{r=r_*}=0$ holds. Passing
$r_*$, $r^{\prime}$ increases as $r$ decreases further, crossing the
$r^{\prime}=0$ as shown in Fig.~\ref{fig:lifetime}.

First, as in the case of $\omega/\mathcal{Q}(\omega)$, two
singularities exist in $\mathcal{Q}(r)-r$. For $r$ close to $r_*$,
Eq.~(\ref{eq:qr}) is expanded as
\begin{equation}
r^{\prime}\simeq r^{\prime}_* +\sum_{n=2}^\infty
\frac{G_n(\gamma)}{n!}(r_*-r)^n, \label{eq:ltexpand1}
\end{equation}
where $r^{\prime}_* =\mathcal{Q}(r_*)-r_* < 0$ and $G_n(\gamma)$ is
the $n$-th derivative of $\mathcal{Q}(r)-r$ at $r^*$. When $r$ is
close to $r_*$ such that
\begin{equation} {\rm {max}}_{n\ge 2}\frac{
G_{n+1}(\gamma)}{G_{n}(\gamma)(n+1)}(r_*-r)\ll 1,\label{condition2}
\end{equation} one may neglect higher order terms, keeping only the
quadratic term in $r_*-r$ as
\begin{equation}
\frac{dr}{dt} \approx r^{\prime}_* + \frac{G_2(\gamma)}{2}
(r_*-r)^2.
\end{equation}
The solution to the above differential equation is
\begin{equation}
r(t)\simeq r(\infty)-\frac{2a}{e^{t/t_*}-1}, \label{eq:rt1}
\end{equation}
where $r(\infty) = r_*-a$ and $a=\sqrt{{2
|r^{\prime}_*|}/{G_2(\gamma)}}$, and $t_*=1/\sqrt{2|r^{\prime}_*|
G_2(\gamma)}$. The lifetime distribution $\ell(t)=r^{\prime}(t)$ is
then given by
\begin{equation}
\ell(t) \simeq  \frac{2ae^{t/t_*}}{t_*(e^{t/t_*} - 1)^2 }\sim
\frac{2a}{t_*}e^{-t/t_*}. \label{eq:lt1}
\end{equation}

Second, following the same steps taken for the singularities of
$\mathcal{P}(y)$, we find another approximate relation between
$r^{\prime}$ and $r$ in the region of $r(t)$ where the next order
term in Eq.~(\ref{eq:qr}) is much larger than its linear term as
follows:
\begin{equation}
\frac{dr}{dt}\sim  \left\{
\begin{array}{ll}
\frac{B(\gamma)}{2}(1-r)^2 & {\rm for}~~~ \gamma>3,\\[2.5mm]
-\frac{1}{2} (1-r)^2 \ln (1-r) & {\rm for}~~~ \gamma=3,\\[2.5mm]
A(\gamma)(1-r)^{\gamma-1} & {\rm for}~~~ 2<\gamma<3.
\end{array}\right.
\label{eq:ltexpand2}
\end{equation}
Their solutions are, in long time limit, given by
\begin{equation}
1-r(t) \sim \left\{
\begin{array}{ll}
t^{-1} & {\rm for}~~~ \gamma>3,\\[2.5mm]
t^{-1}(\ln t)^{-1} & {\rm for}~~~ \gamma=3,\\[2.5mm]
t^{-1/(\gamma-2)} & {\rm for}~~~ 2<\gamma<3.
\end{array}
\right. \label{eq:rt2}
\end{equation}
From these results, the lifetime distributions are obtained as
\begin{equation}
\ell(t) \sim \left\{
\begin{array}{ll}
t^{-2} & {\rm for}~~~ \gamma>3,\\[2.5mm]
t^{-2}(\ln t)^{-1} & {\rm for}~~~ \gamma=3,\\[2.5mm]
t^{-(\gamma-1)/(\gamma-2)} & {\rm for}~~~ 2<\gamma<3.
\end{array}\right.
\label{eq:lt2}
\end{equation}

Different behaviors of the lifetime distribution shown in
Eqs.~(\ref{eq:lt1}) and (\ref{eq:lt2}) suggest the presence of a
crossover behavior. The characteristic time that distinguishes the
two behaviors for given $\gamma$ can be found by considering the
valid ranges of $t$ for Eqs.~(\ref{eq:lt1}) and (\ref{eq:lt2}),
respectively. When the condition of Eq.~(\ref{condition2}) is
fulfilled, Eqs.~(\ref{eq:rt1}) and (\ref{eq:lt1}) are valid. The
condition is approximately represented in different form of
$r_*-r\ll 1-r_*$ since $G_n(\gamma)\sim (1-r_*)^{\gamma-1-n}$. From
Eq.~(\ref{eq:qr}), one can find the value of $1-r_*$ for different
$\gamma$'s: $1-r_*\sim \Delta$ for $\gamma>3$,
$\Delta/\ln(1/\Delta)$ for $\gamma=3$, and $\Delta^{1/(\gamma-2)}$
for $2<\gamma<3$, respectively. Applying these conditions to
Eq.~(\ref{eq:rt1}), it is found that Eqs.~(\ref{eq:rt1}) and
(\ref{eq:lt1}) are valid if $t\gg t_{*1}$ with $t_{*1}\sim
\Delta^{-1}$ irrespective of $\gamma$ as long as $\gamma>2$.

\begin{figure}[t]
\includegraphics[width=8cm]{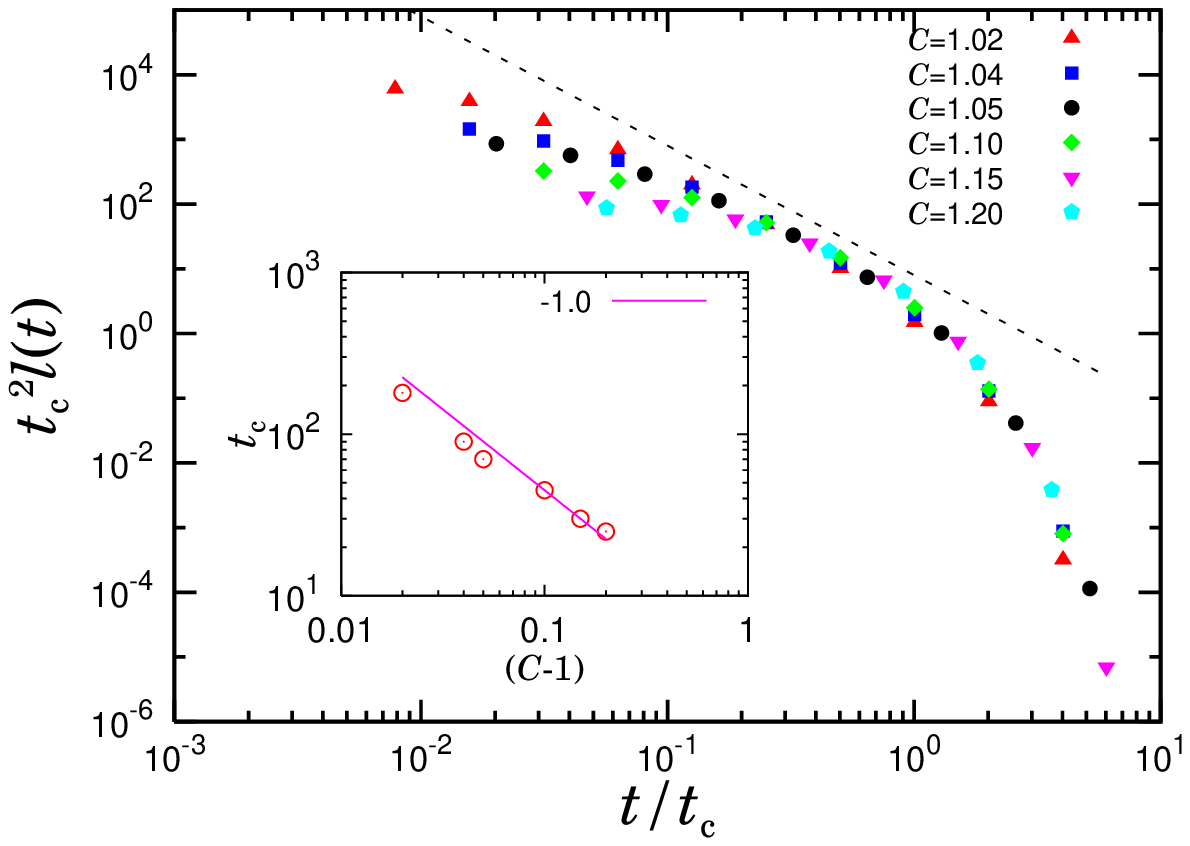}
\caption{(Color online) The lifetime distribution $\ell(t)$ for
$\gamma=3.3$ in the scaling form Eqs.~(\ref{eq:lt3}) and
(\ref{eq:lt4}). Dashed line is guideline with slope $-2$. Data for
small $t$ are deviated from the data collapse, indicating that our
solution is valid for large $t$ only. Inset: Dependence of the
characteristic time $t_c$ on the mean branching number $C$.}
\label{fig:lt33}
\end{figure}
\begin{figure}
\includegraphics[width=8cm]{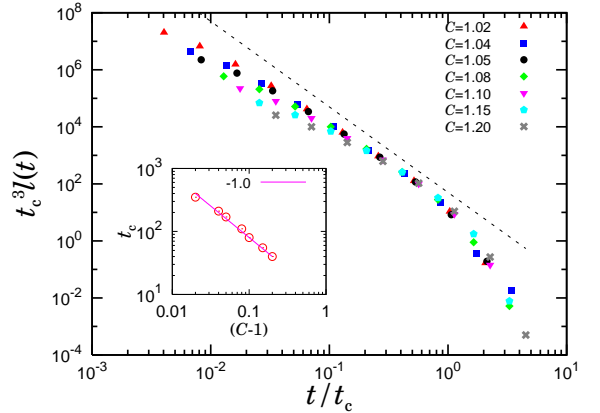}
\caption{(Color online) The lifetime distribution $\ell(t)$ for
$\gamma=2.5$ in the scaling form Eqs.~(\ref{eq:lt3}) and
(\ref{eq:lt4}). Dashed line is guideline with slope $-3$. Data for
small $t$ are deviated from the data collapse, indicating that our
solution is valid for large $t$ only. Inset: Dependence of the
characteristic time $t_c$ on the mean branching number $C$.}
\label{fig:lt25}
\end{figure}

Eqs.~(\ref{eq:rt2}) and (\ref{eq:lt2}) are valid when the linear
term is much smaller than the next order term, which is satisfied
when $1-r\gg \Delta$ for $\gamma>3$, $1-r\gg \Delta/\ln(1/\Delta)$
for $\gamma=3$, and $1-r\gg \Delta^{1/(\gamma-2)}$ for $2<\gamma<3$,
respectively. Applying these conditions to Eq.~(\ref{eq:rt2}) leads
commonly to $t\ll t_{*2}\sim \Delta^{-1}$. One can find that the two
characteristic times $t_{*1}$ and $t_{*2}$, and $t_*$ scale in the
same manner, so that they are denoted as $t_c$ commonly. Therefore,
we conclude that the lifetime distribution behaves as
\begin{equation}
\ell(t)\sim \left\{
\begin{array}{ll}
t^{-2} & {\rm for}~~~ \gamma>3,\\[2.5mm]
t^{-2}(\ln t)^{-1} & {\rm for}~~~ \gamma=3,\\[2.5mm]
t^{-(\gamma-1)/(\gamma-2)} & {\rm for}~~~ 2<\gamma<3,
\end{array}\right.\label{eq:lt3}
\end{equation}
when $t\ll t_c\sim \Delta^{-1}$, and
\begin{equation}
\ell(t)\sim e^{-t/t_c}~~~~~~~ {\rm for}~~~\gamma > 2 \label{eq:lt4}
\end{equation}
when $t \gg t_c$. The analytic solutions for the lifetime
distribution are checked by numerical simulations in
Figs.~\ref{fig:lt33} and \ref{fig:lt25}. Data in small $t$ regime
are somewhat deviated from the data-collapsed formula, indicating
that our solution is valid in large $t$ regime.

\section{Conclusions and Discussion}
Our main results are Eqs.~(\ref{eq:ps_gamma>3}),
(\ref{eq:ps_gamma=3}), and (\ref{eq:ps_2<gamma<3}) for the tree-size
distribution when trees are finite: Contrary to the case of
$\gamma>3$ for which the tree-size distribution $p(s)$ behaves as
$\sim s^{-3/2} \exp(-s/s_c)$ for all $s$ with $s_c \sim (C-1)^{-2}$,
a crossover behavior occurs at $s_c\sim
(C-1)^{-(\gamma-1)/(\gamma-2)}$ for $2<\gamma<3$. For $s \ll s_c$,
$p(s)\sim s^{-\gamma/(\gamma-1)}$ and for $s \gg s_c$, $p(s)\sim
s^{-3/2}\exp(-s/s_c)$. This result is complementary to the previous
mean-field solution $p_{\rm inf}(s)\sim s^{-2}$ for infinite-size
tree. From our solutions, it is noteworthy that the characteristic
size $s_c$ increases as the exponent $\gamma$ approaches 2. This
leads to an interesting result: A larger-size tree can be generated
for smaller value of the exponent $\gamma$. However, the probability
to have such a large-size tree becomes smaller as the exponent
$\gamma$ approaches 2, because the exponent $\gamma/(\gamma-1)$ for
the tree-size distribution $p(s)$ becomes larger.

The lifetime distribution also exhibits a crossover behavior at
$t_c\sim (C-1)^{-1}$. It follows Eq.~(\ref{eq:lt3}) for $t\ll t_c$
and (\ref{eq:lt4}) for $t\gg t_c$.\\

This work was supported by KRF Grant No.~R14-2002-059-010000-0 of
the ABRL program funded by the Korean government (MOEHRD). Notre
Dame's Center for Complex Networks kindly acknowledges the support
of the National Science Foundation under Grant No. ITR DMR-0426737.

\end{document}